\documentclass[pre,twocolumn,superscriptaddress]{revtex4-1}
\usepackage{graphicx}
\usepackage{amsmath,amssymb}

\newcommand{\kB}{k_\mathrm{B}}
\newcommand{\kT}{\kB T}
\newcommand{\lB}{l_{\mathrm{B}}}
\newcommand{\kD}{k_{\mathrm{D}}}
\newcommand{\qD}{q_{\mathrm{D}}}

\newcommand{\fid}{f^{\mathrm{id}}}
\newcommand{\fex}{f^{\mathrm{ex}}}

\newcommand{\latin}[1]{{\itshape #1}}

\newcommand{\ie}{\latin{i.\,e.}}
\newcommand{\etal}{\latin{et al.}}
\newcommand{\via}{\latin{via}}

\newcommand{\half}{\frac{1}{2}}
\newcommand{\quarter}{\frac{1}{4}}

\DeclareMathOperator{\erf}{erf}

\begin{document}

\title{Phase behaviour and the random phase approximation\\ for
  ultrasoft restricted primitive models}

\author{Patrick B. Warren}

\email{patrick.warren@unilever.com}

\affiliation{Unilever R\&D Port Sunlight, Quarry Road East, Bebington,
  Wirral, CH63 3JW, UK.}

\author{Andrew J. Masters}

\affiliation{School of Chemical Engineering and Analytical Science,
  University of Manchester, Manchester M13 9PL, UK.}

\date{February 21, 2013}

\begin{abstract}
Phase separation of the ultrasoft restricted primitive model (URPM) 
with Gaussian charges is re-investigated in the random phase 
approximation (RPA)---the `Level A' approximation discussed by 
Nikoubashman, Hansen and Kahl [J.~Chem.~Phys.~{\bf 137}, 094905 (2012)].  
We find that the RPA predicts a region of low temperature
vapour-liquid coexistence, with a 
critical density much lower than that observed in either simulations 
or more refined approximations (we also remark that the RPA critical 
point for a related model with Bessel charges can be solved 
analytically).  This observation suggests that the hierarchy of 
approximations introduced by Nikoubashman \etal\ should be analogous 
to those introduced by Fisher and Levin for the restricted primitive 
model [Phys.~Rev.~Lett.~{\bf71}, 3826 (1993)], which makes the 
inability of these approximations to capture the observed URPM phase 
behaviour even more worthy of investigation.
\end{abstract}


\maketitle

Recently Coslovich, Hansen and Kahl (CHK) introduced a novel class of
Gaussian charge cloud models for mixtures of interpenetrable
polycations and polyanions in solution \cite{CHK11a, CHK11b}. The low
temperature phase behaviour of these models was explored both by
Monte-Carlo and molecular dynamics simulations \cite{CHK11a, CHK11b},
and in mean field theory by Nikoubashman, Hansen and Kahl
(NHK) \cite{NHK12}. Our interest in this class of models stems from a
different perspective.  In mesoscale models, particularly in
dissipative particle dynamics (DPD) \cite{FS02} soft interactions are
the norm.  Then it is both natural, and indeed essential, to smear out
point charges into charge clouds.  The divergence of the long-range
Coulomb law as $r\to0$ (where $r$ is the center-center separation) is
replaced by a smooth cutoff, thus ensuring thermodynamic stability
according to a theorem by Fisher and Ruelle \cite{MEF66}. The precise
form of the charge smearing is often tuned to the numerical algorithm
used to calculate the electrostatic interactions, and a consensus on
the best approach has yet to emerge \cite{Gro03, GMV+06}.  Whilst for
mesoscale modelling applications the low temperature phase behaviour
is not in itself of primary importance, the screening properties
though are of great interest and our research into this aspect will be
reported more thoroughly elsewhere.

The canonical example of this class of models, which CHK termed the
ultrasoft restricted primitive model (URPM), is an equimolar mixture
of Gaussian charge clouds, which are identical apart from the sign of
the charges, and for which only the electrostatic interactions are
kept.  The URPM is a natural counterpart to the well-studied
restricted primitive model (RPM) of equi-sized charged hard
spheres \cite{Fis94, Ste96, Lev02} in which the short-range Coulombic
divergence is hidden behind the hard core repulsion.  For the URPM,
CHK reported a region of low temperature vapour-liquid phase
coexistence, for which the terminus on increasing temperature bears
many of the hallmarks of a tricritical point.  Above this point, and a
likely reason for the apparent tricriticality, is either a weak second
order transition or a rapid crossover between an insulating dielectric
phase of neutral `molecules' of paired opposite charges and a
conducting `plasma' phase containing a substantial fraction of free
ions.  Subsequently NHK investigated a hierarchy of mean-field
approximations in an attempt to understand in detail the origin of the
low temperature URPM phase behaviour.  This hierarchy was built in
analogy to the earlier work of Fisher and Levin on the RPM \cite{FL93,
  LF96}. The simplest level of approximation, termed `Level A' in NHK,
is analogous to the Fisher-Levin DH (Debye-H\"uckel) approximation.
It is identical to the random phase approximation (RPA) from integral
equation theory, and incorporates the mutual attractions and
repulsions in a linearised way.  The next level of approximation,
`Level B' in Ref.~\cite{NHK12} and DHBj (Debye-H\"uckel-Bjerrum)
in the Fisher-Levin classification, captures the formation of ion
pairs---a crucial aspect of the non-linear physics at low
temperatures.

NHK assert that ``there is no phase separation at [the `Level A']
approximation'' (below Eq.~(30) in Ref.~\cite{NHK12}).  This caught
our attention, as we have known for some time that the RPA for a
related Bessel charge model (discussed below) 
does exhibit phase separation, with a critical point which can be
determined analytically.  Prompted by this discrepancy, our own
further investigations reveal that the RPA for Gaussian charges
\emph{does} have a region of phase separation, but at a much lower
density than investigated by NHK.

To set the problem up, we consider an equimolar mixture of 
$N_+=N_-=N/2$ charge clouds (polyions) in a volume $V$, with an
overall density $\rho=N/V$.  Gaussian charge clouds interact with the
following pair potential,
\begin{equation}
u_{\pm\pm}(r)=\pm u(r),\quad \beta
u(r)=\frac{\lB}{r}\erf\Bigl(\frac{r}{2\sigma}\Bigr)
\label{eq:gurpm}
\end{equation}
where $u(r)$ is the pair potential between charge clouds of the same
sign, $\beta=1/\kT$ is the inverse of the temperature measured in
units of Boltzmann's constant, $\lB$ is the Bjerrum length which plays
the role of a coupling constant, $r$ is the separation, and $\sigma$
is a measure of the size of the charge cloud.  For Gaussian charges
the radial charge distribution corresponding to this potential is
$(2\pi\sigma^2)^{-3/2} e^{-r^2/2\sigma^2}$.  The function
$\erf(r/2\sigma) \sim r$ as $r\to0$, thus ensuring the Coulombic
divergence is replaced by a smooth cutoff.  

An interesting alternative to the Gaussian charge URPM is provided by
a Bessel charge model.  For this case the interaction potential is simply
\begin{equation}
\beta u(r)=\frac{\lB}{r}(1-e^{-r/\sigma})
\label{eq:burpm}
\end{equation}
This corresponds to a radial charge distribution $K_1(r/\sigma) /
(2\pi^2\sigma^2 r)$ where $K_1$ is a modified Bessel function (hence
the name).  Although this radial charge distribution diverges as
$r\to0$, the interaction potential itself is again smoothly cutoff.

Gaussian charges are blessed by being particularly well suited to the
Ewald summation method for handling long range Coulomb interactions,
as has been noted by CHK.  Bessel charges, on the other hand, are not
so well suited for simulations but provide perhaps the simplest
non-trivial example of an ultrasoft primitive model when it comes to
analytical work.  Obviously, the definition of $\sigma$ in the two
potentials cannot be exactly matched up and this should be born in
mind when making comparisons.

In reciprocal space these potentials are
\begin{equation}
\beta\tilde u(k)=\frac{4\pi\lB}{k^2}\,w(k\sigma)
\end{equation}
where, writing $q=k\sigma$,
\begin{equation}
w(q)=\left\{\begin{array}{ll}
e^{-q^2} & \text{(Gaussian),}\\[3pt]
1/(1+q^2) & \text{(Bessel).}
\end{array}\right.
\end{equation}
The definition of $\sigma$ in the two models is chosen to
match up the long wavelength behaviour here.

\begin{figure}
\begin{center}
\includegraphics[clip=true,width=2.8in]{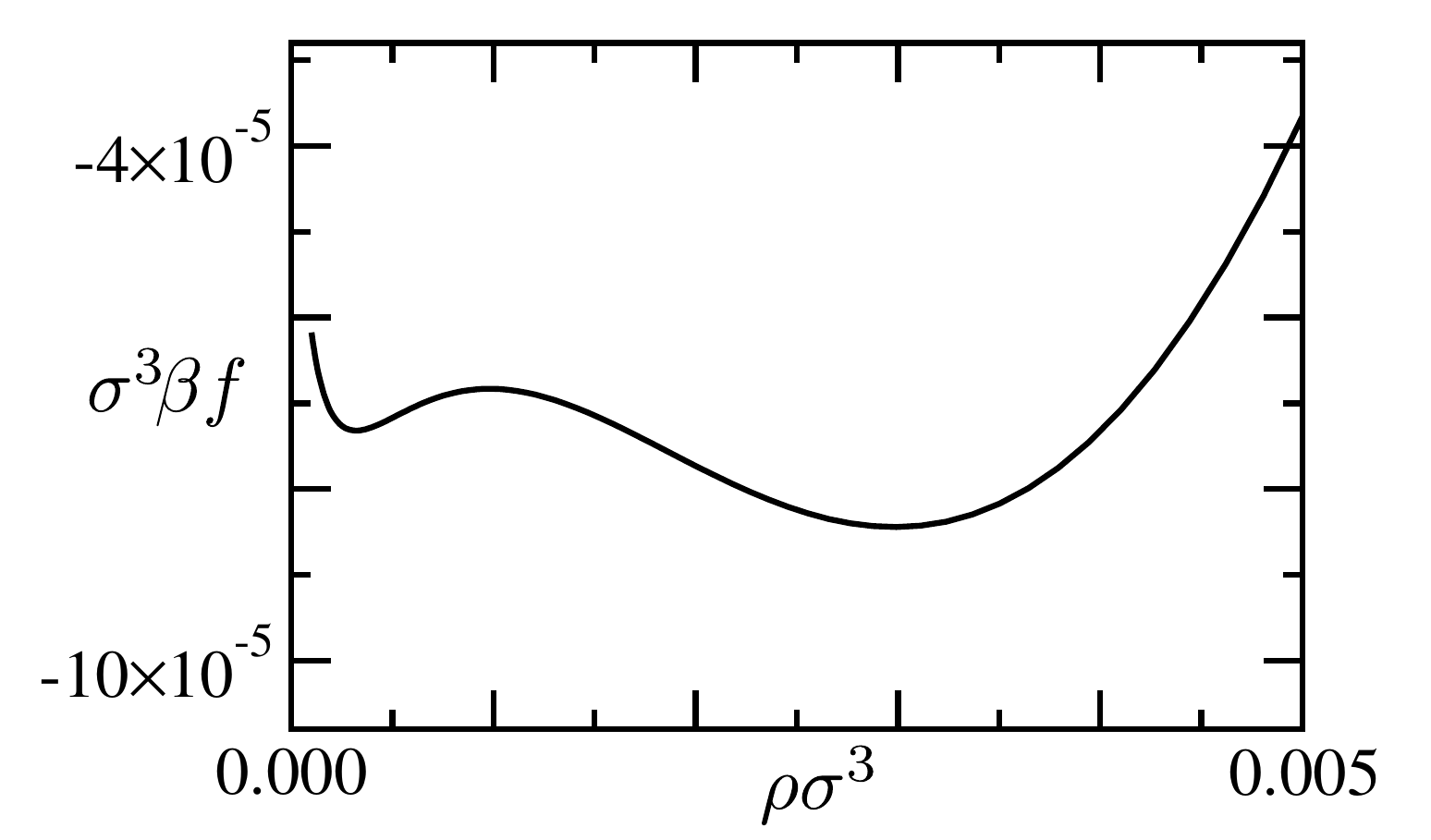}
\end{center}
\vskip -0.5cm
\caption{RPA free energy for Gaussian URPM at
  $\sigma\sqrt{\pi}/\lB=\sqrt{\pi}/27\approx0.0656$, from
  Eqs.~\eqref{eq:fid}--\eqref{eq:ftot} with $w(q)=e^{-q^2}$.  A
  function $A\rho$, with $\beta A=4.03$, is added to the free energy
  to reveal the common tangent construction without perturbing the
  phase behaviour.\label{fig:rpaf}}
\end{figure}

The random phase approximation (RPA) for this class of models takes
the form $c_{\pm\pm}=-\beta u_{\pm\pm}$ for the direct correlation
functions \cite{FHL03, HAE06, CHK11b, NHK12, HM06}.  Because of the absence
of hard cores, the RPA is also equivalent to the mean spherical
approximation (MSA).  From the RPA, the total correlation functions,
$h_{\pm\pm}(r)=\pm h(r)$, follow by inversion of the Ornstein-Zernike
equations.  In reciprocal space the solution is
\begin{equation}
\tilde h(k)=\frac{-4\pi\lB w(k\sigma)}{k^2+\kD^2 w(k\sigma)}\,.
\label{eq:hk}
\end{equation}
In this $\kD^2\equiv 4\pi\lB\rho$ is the square of the Debye
wavevector.  It follows from Eq.~\eqref{eq:hk} that the
density-density structure factor is given by $S_{NN}(k)=1$ and,
somewhat less trivially, the charge-charge structure factor is given
by $S_{ZZ}(k)=k^2/[k^2+\kD^2 w(k\sigma)]$.

In all these we notice the prominent role played by the denominator
$D(k)=k^2+\kD^2w(k\sigma)$.  As is well known \cite{HAE06, NHK12} the
zeros of this function in the complex $k$-plane determine the
asymptotic behaviour of the total correlation functions, and are
crucial to understanding the screening properties of the system
particularly for applications in mesoscale modelling.  The asymptotic
behaviour typically crosses over from being purely exponential to
being damped oscillatory as one increases the density past the
so-called Kirkwood line in the density-temperature plane \cite{Kir36}.
More generally, this is referred to as a Fisher-Widom line \cite{FW69}.
For Gaussian charges the asymptotic behaviour is determined by the
complex roots of $q^2 + \qD^2 e^{-q^2}\! = 0$, where $\qD =
\kD\sigma$.  The most relevant roots are given by $q^2=W_0(-\qD^2)$
where $W_0$ is the principal branch of the Lambert $W$
function \cite{Cor96}.  From this, or by direct
calculations \cite{NHK12}, the Kirkwood line for Gaussian charges is
given by $\qD=e^{-1/2}\approx 0.6065$.  The Kirkwood line for Bessel
charges is determined by the complex roots of the biquadratic equation
$q^4 + q^2 + \qD^2 = 0$.  For $\qD\le\half$ the roots are all purely
imaginary, whereas for $\qD>\half$ they are all complex.  Hence in
this case the Kirkwood line takes the simple form $\qD=\half$.

\begin{figure}
\begin{center}
\includegraphics[clip=true,width=2.8in]{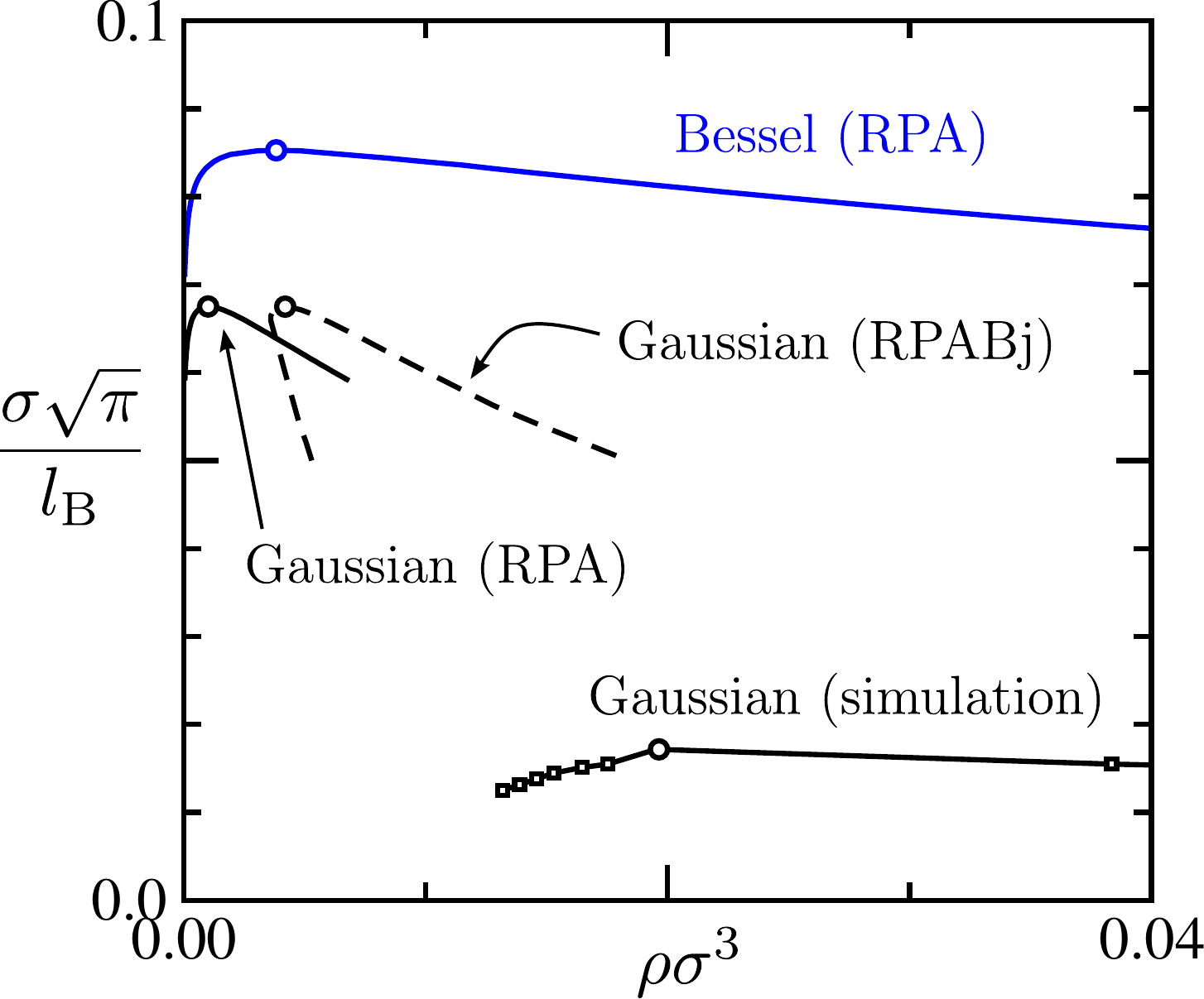}
\end{center}
\vskip -0.5cm
\caption{(color online) Vapour-liquid coexistence regions (binodals
  plus critical points) for the URPM with Gaussian or Bessel charges.
  Approximations are RPA (present work) and RPABj (`Level B' in
  Ref.~\cite{NHK12}).  Data for RPABj is taken from Fig.~7 of
  Ref.~\cite{NHK12}, and the simulation data is taken from
  Fig.~18 of Ref.~\cite{CHK11b}.  See Table \ref{tab:cp} for
  locations of critical points.\label{fig:rpaphase}}
\end{figure}

Now we turn to the free energy.  It follows from the density-density
structure factor that the compressibility-route equation of
state is trivially that of an ideal gas, for which the free energy
density is 
\begin{equation}
\beta\fid=\rho(\ln{{\textstyle\half}\rho}-1)\,.
\label{eq:fid}
\end{equation}
Note there are two species of ions contributing to this, each at a
density $\half\rho$, and we have neglected the thermal de Broglie
wavelength as it plays no role in phase coexistence.  The virial-route
equation of state, and the energy-route equation of state
(\via\ coupling constant integration) give rise to the same result,
which can be integrated to a non-trivial excess free energy density.
The result is
\begin{equation}
\beta\fex=\frac{1}{4\pi^2\sigma^3}\int_0^\infty \!\!dq\,
\Bigl[q^2\ln\Bigl(1+\frac{\qD^2}{q^2}w(q)\Bigr)-\qD^2w(q)\Bigr]\,.
\label{eq:fex}
\end{equation}
For point charges $w(q)=1$ and this reduces to the exact DH limiting
law $\beta\fex=-\kappa^3/12\pi$.  The total free energy density, used
in calculating the phase behaviour, is given by the sum of
Eqs.~\eqref{eq:fid} and \eqref{eq:fex} :
\begin{equation}
f=\fid+\fex\,.\label{eq:ftot}
\end{equation}
In the Gaussian case Eq.~\eqref{eq:fex} is exactly equal to Eq.~(29)
in Ref.~\cite{NHK12}.  Figure \ref{fig:rpaf} shows the total
free energy, from Eq.~\eqref{eq:ftot}, as a function of density at a
judiciously chosen temperature, illustrating the existence of a common
tangent construction.  The full phase behaviour is plotted in
Fig.~\ref{fig:rpaphase}, marked `RPA', where also are shown the `Level
B' results replotted from Ref.~\cite{NHK12}, here marked
`RPABj', and simulation results taken from Ref.~\cite{CHK11b}.
The Gaussian RPA critical point, found numerically, is located at
$\lB/\sigma\approx 26.25$ and $\rho\sigma^3\approx1.014\times10^{-3}$
(see also Table \ref{tab:cp}).  This corresponds to a reduced Debye
wavelength of $\qD\approx 0.335$ which places the critical point
somewhat on the low density side of the RPA Kirkwood line.

At this point we should comment on the choice of reduced
(dimensionless) temperature.  CHK and NHK use $u_0=u(0)$ as an energy
scale but this frustrates direct comparison with the RPM.  Our own
preference is to use the \emph{long} range behaviour of the potential
characterised by the reduced Bjerrum length $\lB/\sigma$.  Since
$\beta u_0 = \lB / \sigma\sqrt{\pi}$ for Gaussian charges, to
facilitate the comparison with CHK and NHK we universally use
$\sigma\sqrt{\pi}/\lB$ as a reduced temperature.  In this, $\sigma$ is
the parameter entering the interaction potentials in
Eqs.~\eqref{eq:gurpm} and~\eqref{eq:burpm} for the URPM, and the hard
sphere diameter for the RPM.

\begin{table}
\begin{ruledtabular}
\begin{tabular}{lllll}
system & method & $\sigma\sqrt{\pi}/\lB$ & $\rho\sigma^3$ & Refs.\\[2pt]
\hline
RPM     & DH          & 0.11  & 0.005\,0 & \cite{FL93} \\ 
        & DHBj        & 0.11  & 0.045    & \cite{FL93} \\
        & simulation  & 0.089 & 0.080    & \cite{LFP02}, \cite{CLW02} \\
\hline
URPM (G) & RPA         & 0.068 & 0.001\,0 &  \\
        & RPABj       & 0.068 & 0.004\,2 & \cite{NHK12} \\
        & simulation  & 0.018 & 0.020    & \cite{CHK11a}, \cite{CHK11b} \\
\hline
URPM (B) & RPA         & 0.085 & 0.003\,8 & \\
\end{tabular}
\end{ruledtabular}
\caption[?]{Vapour-liquid critical points for the RPM and URPM with
  Gaussian (G) and Bessel (B) charges.  RPA results are those reported
  in the present paper.  For the Gaussian URPM the RPABj (`Level B' in
  Ref.~\cite{NHK12}) and simulation results are taken from Table
  I in Ref.~\cite{NHK12}.  All results (both here and in the
  main text) are accurate to the final digit.\label{tab:cp}}
\end{table}

For the Bessel case, the RPA excess free energy can be obtained in
closed form.  The last term in Eq.~\eqref{eq:fex} evaluates to
$-\qD^2/(8\pi\sigma^3)=-\lB\rho/(2\sigma)$.  On multiplying through by
$\sigma^3$, the first part of the integral is
\begin{equation}
I=\frac{1}{4\pi^2}\int_0^\infty \!\!dq\,
q^2\ln\Bigl(1+\frac{\qD^2}{q^2(1+q^2)}\Bigr)\,.
\end{equation}
To solve this, we learn from the (Schwinger-)Feynman parameter
trick \cite{RPF49} and rewrite it as
\begin{equation}
I=\frac{1}{4\pi^2}\int_0^\infty \!\!dq
\int_0^{\qD^2}\!\!du\,
\frac{q^2}{q^2(1+q^2)+u}\,.
\end{equation}
Making for the time being the assumption that $\qD^2\le\quarter$ (so
that we are on the low density side of the Kirkwood line), the
$q$-integral can now be done, by the method of partial fractions, to get
\begin{equation}
I=\frac{1}{8\pi\sqrt{2}}
\int_0^{\qD^2}\!\!du\,
\frac{\sqrt{1+z}-\sqrt{1-z}}{z}
\end{equation}
where $z=\sqrt{1-4u}$ (hence the temporary restriction on $\qD$).  We
note that $du=-\half z\,dz$, so the $u$-integral can also be done.
After taking careful account of the integration limits, the final
result for the free energy is
\begin{equation}
\sigma^3\beta\fex=
\frac{2\sqrt{2}-({1+z})^{3/2}-({1-z})^{3/2}}{24\pi\sqrt{2}}
-\frac{\lB\rho\sigma^2}{2}
\label{eq:fexb}
\end{equation}
where now $z=\sqrt{1-4\qD^2}$.  Whilst this result has been derived
for $\qD^2\le\quarter$, it holds by analytic continuation for all $\qD$.  As
one crosses the Kirkwood line from low to high density, $z$ crosses
over from being purely real to purely imaginary, so that
\begin{equation}
z=\left\{\begin{array}{ll}
\sqrt{1-4\qD^2} & (\qD\le\half)\\[3pt]
i\sqrt{4\qD^2-1} & (\qD>\half)\end{array}\right.
\end{equation}
Nevertheless the free energy remains purely real and is continuous
across the Kirkwood line. (Note that the roots of $q^4 + q^2 + \qD^2 =
0$ are given by $q^2=-\half\pm\half z$.)

Like the Gaussian case, the RPA free energy for the Bessel case has a
region of vapour-liquid phase coexistence at low densities and
temperatures.  The critical point can be found by solving $\partial^2
\!f/\partial\rho^2=\partial^3\!f/\partial\rho^3=0$ from
Eqs.~\eqref{eq:fid}, \eqref{eq:ftot} and \eqref{eq:fexb}.  An analytic
solution can be obtained, which is $\lB/\sigma=12\sqrt{3}\approx20.78$
and $\rho\sigma^3=1/(48\pi\sqrt{3})\approx3.829\times10^{-3}$ (see
also Table \ref{tab:cp}).  This corresponds to $z=i\sqrt{3}$ and
$\qD=1$, thus for Bessel charges the RPA critical point lies on the
high density side of the Kirkwood line.  The phase behaviour for the
Bessel case, calculated numerically, is also shown in
Fig.~\ref{fig:rpaphase}.

Table \ref{tab:cp} compares the vapour-liquid critical points for the
RPM and the URPM, using various approximations.  We see that the DH
approximation for the RPM, and the RPA for the URPM, both predict
critical points at low densities and temperatures.  When Bjerrum
pairing is incorporated (\ie\ DHBj for RPM, and RPABj for Gaussian
URPM), the critical temperature remains unchanged but the critical
density is considerably increased.  For the RPM, this brings the
predicted critical point quite close to the simulations, within 20\%
for the critical temperature (for a detailed discussion, see
Ref.~\cite{FL93}).  For the URPM with Gaussian charges though,
the predicted critical point is still considerably distant from the
simulations.  In particular the predicted critical temperature is at
least a factor of three above the observed value.  We can to some
extent confirm this observation, as we have looked for phase
separation in the Gaussian URPM using Monte-Carlo methods, at
temperatures in the vicinity of the RPA critical point, and
have found no evidence of such.  This singular aspect of the phase
behaviour of the URPM stands in marked contrast to the RPM.  Some
possible explanations have been proposed by NHK \cite{NHK12}.

The observation that the critical temperature remains unchanged in
comparing RPA and RPABj can be traced to the fact that in the latter
approximation the Bjerrum pairs are an ideal spectator
species \cite{FL93, LF96, Lev02}.  As such they cannot, in themselves,
influence the phase behaviour of the unpaired ions.  The
quasi-chemical equilibrium between paired and unpaired ions changes
the coexistence densities, in accordance with the law of mass action,
but the critical temperature itself remains unaffected.

To summarise, the Fisher-Levin hierarchy of approximations developed
for the RPM can be pursued also for the URPM, with similar trends,
indicating the two models \emph{should} show similar phase behaviour.
The fact that they do not deepens the mystery uncovered by
Nikoubashman, Hansen and Kahl in Ref.~\cite{NHK12} and clearly
warrants further investigation.


%

\end{document}